\documentclass[twocolumn,showpacs,preprintnumbers,amsmath,amssymb]{revtex4}

\usepackage{graphicx}% Include figure files
\usepackage{dcolumn}% Align table columns on decimal point
\usepackage{bm}% bold math
\usepackage{amsmath,amssymb,amsfonts}
\makeatletter

\newcommand{\Rmnum}[1]{\expandafter\@slowromancap\romannumeral #1@}
\makeatother

\begin{document}

\title{Non-Abrikosov vortices in liquid metallic hydrogen }
\author{Liping Zou}
\email{zoulp@impcas.ac.cn}
\affiliation{%
Institute of Modern Physics, Chinese Academy of Sciences, Lanzhou 730000, China
\\
School of Nuclear Science and Technology, Lanzhou University, Lanzhou 730000, China
\\
University of Chinese Academy of Sciences, Beijing 100049, China
}%

\begin{abstract}
We consider non-Abrikosov vortex solutions in liquid metallic hydrogen (LMH)
in the framework of two-component Ginzburg-Landau model.
We have shown that there are
three types of non-Abrikosov vortices depending on chosen
boundary conditions at the core of vortices, namely, Neumann (N)-type, Dirichlet (D)-type
and Gross-Pitaevskii (GP)-type vortices. The
Neumann-type vortex has a non-vanishing condensation
at the core, that is different from the ordinary vortex, and the magnetic
flux could be reversed as well in LMH. Furthermore, we have obtained
a new type of a neutral vortex which has no magnetic field. The presence of such a vortex
is related to metallic superfluid state suggested by Babaev\cite{Babae04}.
\end{abstract}


\maketitle

\section{Introduction }
The Abrikosov vortex has been intensively studied in one-component superconductor.
Recently, the interest in non-Abrikosov vortices in two-component or multi-component superconductor
has been increased widely due to their unusual properties\cite{Babae04,Cho08,Juan2012,Babaev05,Geurt2010,Chaves2011,Babaev101,Babaev102,Johan1,Johan2,Moshchalkov09}.
The non-Abrikosov vortex may have a fractional magnetic flux
and non-vanishing condensation at the core of the vortex\cite{Babae04,Cho08,Juan2012,Babaev02}.
In comparison with the Abrikosov vortex, the non-Abrikosov vortex may possesses
a specific kind of interaction which is repulsive at short distances and
attractive at larger scales\cite{Babaev05}, details for for nonmonotonic vortex interaction in two-band superconductors
were discussed in\cite{Chaves2011}. Besides, new interesting phenomena, like the presence of
type-1.5 superconductivity and an abnormal external field response, have been found
in two-component superconductor \cite{Babaev05,Babaev101,Babaev102,Johan1,Johan2,Babaev2013,Moshchalkov09}.
A liquid metallic hydrogen represents one of possible realizations of two-component superconductor
medium where the existence of non-Abrikosov vortices and their characteristics
can be verified in experiment. Another attractive feature of the LMH
is that it could be an alternative candidate for the high temperature superconductor
with coexistence phase of electron-electron and proton-proton Cooper pairs\cite{Ashcroft68,Moulopolos91}.
Such a multi-component superconductor dresses novel features that do not appear
in the normal superconductor.

In the Ginzburg-Landau (GL) theory of superconductivity,
the GL parameter $\kappa$ defined as a
ratio of the penetration length $\lambda$ to the coherence length $\xi$
divides the superconductor into two classes, type-I and type-II superconductor
with the parameter values $\kappa<1/\sqrt{2}$ and $\kappa>1/\sqrt{2}$ respectively.
Since the two-component superconductor has two independent GL parameters, $\kappa_{1,2}$,
a new type of superconductivity appears in the regime $\kappa_1<1/\sqrt{2}$
and $\kappa_2>1/\sqrt{2}$ (assuming $\kappa_1 < \kappa_2$)
which is coined as a type-1.5 superconductivity  \cite{Babaev05,Moshchalkov09}.
Another interesting issue is that there are different types of topologically stable
non-Abrikosov vortex solutions in LMH. Except the D-type and N-type vortices discussed in \cite{Cho08},
a new GP-type vortex has been found in LMH.
A main purpose of the present paper is to study these new vortex solutions and their
properties in LMH. It has been shown that type-II and type-1.5
superconductivity may exist in LMH\cite{Babaev101,Babaev102,Johan1,Johan2,Babaev2013,Moshchalkov09},
and for each type of superconductivity there are three
types of non-Abrikosov vortex solutions. We consider the following
properties of the vortices related to the presence of the fractional magnetic
flux, non-vanishing condensate at the core of the vortex,
magnetic flux inversion and a neutral vortex.
An extended GL theory has been studied in \cite{vagov2012} which has advantages in studying of
electronic, magnetic, calorimetric, and other properties of twoband superconductors. And the GL theory for multiband superconductors
from multiband BCS Hamiltonian is discussed in \cite{Orlova2013}.

The paper is organized as follows. In Section 2 we introduce the
two-component Ginzburg-Landau theory for LMH.
In Section 3 we present numerical vortex solutions of LMH with two condensates
having only one effective complex phase factor.
The D-type and N-type vortices with unusual properties
are described as well.
Section 4 deals with the case of both condensates with non-zero complex phases.
We demonstrate that there exist three types of vortex solutions,
D-type, N-type and GP-type vortices, determined by
different boundary conditions at the core.
We have found a neutral vortex in LMH which has no magnetic field.
Conclusions and discussions are given in the last section.

\section{Ginzburg-Landau model }

Let us start with a free energy of LMH described by two-component Ginzburg-Landau
model\cite{Babae04,Cho08,Forgacs06}

\begin{equation}
\begin{aligned}
\mathcal{H}=&\dfrac{\hbar^2}{2m_1}|({\bf \nabla}+ig_1{\bf A})\tilde{\phi}_1|^2+
\dfrac{\hbar^2}{2m_2}|({\bf \nabla}-ig_2{\bf A})\tilde{\phi}_2|^2\\
&
+\dfrac{1}{2}({\bf \nabla}\times {\bf A})^2+V(\tilde{\phi}_1,\tilde{\phi}_2),
\label{eq:Hamiltonian1}
\end{aligned}
\end{equation}
where $~\tilde{\phi}_1~$ and $~\tilde{\phi}_2~$ are two complex fields corresponding
to the order parameters of electron and proton condensates, and $g_1$ and $g_2$
are absolute values of the Cooper pair's charge, $g_1=g_2=2e$, $m_1,~m_2$
are the masses of electronic and protonic Cooper pairs, respectively.

In LMH the both, electron and proton, condensates are conserved independently
since the electronic Cooper pairs cannot convert to protonic Cooper pairs.
Therefore, there is no intrinsic
Josephson interband interaction$~\eta(\phi^\ast_1\phi_2+h.c.)~$in LMH\cite{Babae04}.
Moreover, there is no interband interaction $~\lambda_{12}|\phi_1|^2|\phi_2|^2~$
in two-gap metallic superconductors\cite{Zhitomirsky04}.
In this weak-coupling approximation the effective potential for LMH can be expressed as follows
\begin{equation}
  V(\tilde{\phi}_1,\tilde{\phi}_2)=\dfrac{1}{2}\tilde{\lambda}_{11}|\tilde{\phi}_1|^4+
\dfrac{1}{2}\tilde{\lambda}_{22}|\tilde{\phi}_2|^4-\tilde{\mu}_1|\tilde{\phi}_1|^2-
\tilde{\mu}_2|\tilde{\phi}_2|^2,
\end{equation}
where $\tilde{\lambda}_{11,22}$ are the quartic coupling constants
and $\tilde{\mu}_{1,2}$ are chemical potentials.
One can simplify the Hamiltonian (\ref{eq:Hamiltonian1})
with the normalized valuables $\phi_{1,2}=\hbar\tilde{\phi}_{1,2}/\sqrt{2m_{1,2}}$ \cite{Cho08},
\begin{equation}
\mathcal{H}=|({\bf \nabla}+ig{\bf A})\phi|^2+V(\phi_1,\phi_2)+\dfrac{1}{2}({\bf \nabla}
\times {\bf A})^2,
\label{eq:Hamiltonian2}
\end{equation}
where $\phi=(\phi_1,\phi_2)$ is a complex doublet, and the normalized potential $V$ is given by
\begin{equation}
V(\phi_1,\phi_2)=\dfrac{\lambda_{11}}{2}|\phi_1|^4-\mu_1|\phi_1|^2
+\dfrac{\lambda_{22}}{2}|\phi_2|^4-\mu_2|\phi_2|^2.
\end{equation}
where $\lambda_{11,22}$ are the normalized quartic coupling constants and $\mu_{1,2}$ are the normalized chemical potentials.

In addition, we can rewrite the potential $V(\phi_1,\phi_2)$
in the form
\begin{equation}
\begin{aligned}
V=&\dfrac{\beta}{2}(|\phi_1|^2+|\phi_2|^2-\dfrac{\mu}{\beta})^2
+\dfrac{\alpha}{2}(|\phi_1|^4-|\phi_2|^4)\\
&+\dfrac{\beta}{2}(|\phi_1|^2-|\phi_2|^2)^2-\gamma(|\phi_1|^2-|\phi_2|^2)\\
&-\dfrac{\mu^2}{2\beta},
\end{aligned}
\end{equation}
where $\alpha=(\lambda_{11}-\lambda_{22})/2,~\beta=(\lambda_{11}+\lambda_{22})/4,~\mu=(\mu_1+\mu_2)/2,~\gamma=(\mu_1-\mu_2)/2$.

At space infinity the vacuum expectations of $\phi_1,\phi_2~$ are given by the  London limit
\begin{equation}
\langle |\phi_1|^2\rangle=\dfrac{\mu_1}{\lambda_{11}}=\dfrac{\mu+\gamma}{2\beta+\alpha},~~~~
\langle |\phi_2|^2\rangle=\dfrac{\mu_2}{\lambda_{22}}=\dfrac{\mu-\gamma}{2\beta-\alpha}.
\label{eq:vacuum}
\end{equation}
The masses of the scalar fields and vector field can be expressed
as follows\cite{Hindmarsh92,Nielsen73}
\begin{subequations}
\label{eq:whole}
\begin{equation}
m_{\phi_1}=\sqrt{2\mu_1},
\end{equation}
\begin{equation}
m_{\phi_2}=\sqrt{2\mu_2},
\end{equation}
\begin{equation}
m_{A}=\sqrt{2g^2(\langle|\phi_1|^2\rangle+\langle|\phi_2|^2\rangle)}.
\end{equation}
\end{subequations}
where $m_{\phi_{1,2}}=1/\xi_{1,2}$, $\xi_{1,2}$ is the characteristic
length of $\phi_{1,2}$, $m_{A}=1/\lambda$, $\lambda$ is the penetration length.
It is clear that there are two mass ratios due to the presence of two condensates in LMH
\begin{subequations}
\begin{equation}
\beta_1=\dfrac{m_{\phi_1}}{m_A}=\sqrt{\dfrac{\lambda_{11}\lambda_{22}\mu_1}
{(\lambda_{11}\mu_2+\lambda_{22}\mu_1)g^2}},
\end{equation}
\begin{equation}
\beta_2=\dfrac{m_{\phi_2}}{m_A}=\sqrt{\dfrac{\lambda_{11}\lambda_{22}\mu_2}
{(\lambda_{11}\mu_2+\lambda_{22}\mu_1)g^2}}.
\end{equation}
\end{subequations}
So that LMH is different from the one-component superconductor.
In the case of $\beta_1>1,$~and~$\beta_2>1,$ it represents type-II superconductivity,
while when $\beta_1<1,$ and $~\beta_2<1$
it has type-I superconductivity. A new feature appears in the region $\beta_1<1,~\beta_2>1$~or~$\beta_1>1,~\beta_2<1$,
where a new, type-1.5, superconductivity has been arised\cite{Babaev05,Moshchalkov09}.

A self-dual solution in two-component GL model may also be of interest,
and it is considered in two-band superconductor in\cite{Cho05PRB},
where two condensates have a non-trivial interband interaction $\lambda_{12}\neq 0$.
When $\mu_1=\mu_2,~\beta_1=\beta_2=\lambda/g^2=1$ the condensates satisfy
Bogomol'nyi first-order equations which have self-dual vortices
as solutions. As we mentioned above, in LMH the interband interaction can be neglected.
Due to this the energy minimization does not imply first-order
equations, so that self-dual vortex does not exist in LMH.

\section{\label{sec:level3}Vortex solutions with one-component complex phase }

We consider a straight vortex with translational symmetry along the $z$
direction and rotational symmetry in the $(x,y)$ plane using
two kinds of ansatz \cite{Cho08,Cho05PRB,Horvathy2009}.
First, we choose an ansatz with two condensates having only one effective complex phase
in cylinder coordinates $(r,\varphi,z)$
\begin{subequations}
\label{eq:ansatz1}
\begin{equation}
\phi=\left(
\begin{array}{c}
\phi_1\\
\phi_2
\end{array}
\right)
=\dfrac{\rho(r)}{\sqrt{2}}
\left(
\begin{array}{c}
\cos{\dfrac{f(r)}{2}}\exp{(-in\varphi)}\\
\sin{\dfrac{f(r)}{2}}
\end{array}
\right),
\end{equation}
\begin{equation}
A_{\mu}=\dfrac{n}{g}A(r)\partial_{\mu}\varphi.
\end{equation}
\end{subequations}
Substituting Eq.~(\ref{eq:ansatz1}) into Eq.~(\ref{eq:Hamiltonian2}) yields
the Hamiltonian
\begin{equation}
\begin{aligned}
\mathcal{H}=&\frac{1}{2}\dot{\rho}^2+\dfrac{1}{8}\rho^2(\dot{f}^2+\dfrac{n^2}{r^2}\sin^2f)+
\dfrac{n^2\rho^2}{2r^2}(A-\dfrac{\cos{f}+1}{2})^2\\
&+\dfrac{n^2}{2g^2r^2}\dot{A}^2
+\dfrac{\beta}{8}[(\rho^2-\dfrac{2\mu}{\beta})^2+\dfrac{\alpha}{\beta}(\rho^2-
\dfrac{4\gamma}{\alpha})\rho^2\cos{f}\\
&+
\rho^4\cos ^2f]-\dfrac{\mu^2}{2\beta}.
\end{aligned}
\end{equation}
With this, the equations of motion become
\begin{subequations}
\label{eq:motion1}
\begin{equation}
\begin{aligned}
&\ddot{\rho}+\dfrac{1}{r}\dot{\rho}-[\dfrac{1}{4}(\dot{f}^2+\dfrac{n^2}{r^2}\sin^2{f})
+\dfrac{n^2}{r^2}(A-\dfrac{\cos{f}+1}{2})^2]\rho\\
&=
\dfrac{\beta}{2}[(\rho^2-\dfrac{2\mu}{\beta})
+\dfrac{\alpha}{\beta}(\rho^2-\dfrac{2\gamma}{\alpha})\cos{f}
+\rho^2\cos^2{f}]\rho,
\end{aligned}
\end{equation}
\begin{equation}
\begin{aligned}
&\ddot{f}+(\dfrac{1}{r}+2\dfrac{\dot{\rho}}{\rho})\dot{f}
-2\dfrac{n^2}{r^2}(A-\dfrac{1}{2})\sin{f}\\
&=[2\gamma-(\dfrac{\alpha}{2}+\beta\cos{f})\rho^2]\sin{f},
\end{aligned}
\end{equation}
\begin{equation}
\ddot{A}-\dfrac{1}{r}\dot{A}-g^2\rho^2(A-\dfrac{\cos{f}+1}{2})=0.
\end{equation}
\end{subequations}
The corresponding electromagnetic current reads
\begin{equation}
j_{\mu}=ng\rho^2\left(A-\dfrac{\cos f+1}{2}\right)\partial_{\mu}\phi.
\end{equation}
Boundary conditions at infinity can be fixed by the vacuum expectation of
the order parameters
\begin{equation}
\begin{aligned}
&\rho(\infty)=2\langle|\phi|^2\rangle=2\sqrt{\dfrac{2\beta\mu-\alpha\gamma}{4\beta^2-\alpha^2}},
\\
&\cos{f(\infty)}=\dfrac{2(\langle|\phi_1|^2\rangle-\langle|\phi_2|^2\rangle)}{\rho^2(\infty)}=\dfrac{2\beta\gamma-\alpha\mu}{2\beta\mu-\alpha\gamma}.
\end{aligned}
\end{equation}

In particular, the electromagnetic current vanishes at infinity, i.e. $j_{\mu}=0$,
so that we have
\begin{equation}
A(\infty)=\dfrac{\cos{f(\infty)+1}}{2}=\dfrac{2\beta(\gamma+\mu)
-\alpha(\gamma+\mu)}{2(2\beta\mu-\alpha\gamma)}.
\end{equation}

On the other hand, boundary conditions at the core can be obtained by
substituting Taylor expansions for the functions $~A,~f,~\rho~$ into the equations of motion
and imposing regularity conditions at the core \cite{Cho08}.
D-type vortex is defined by imposing Dirichlet boundary condition for
the total condensate density $\rho(r)$
\begin{equation}
\rho(0)=0,~~~~A(0)=-\frac{1}{n},~~~~f(0)=\pi.
\end{equation}
N-type vortex corresponds to imposed Neumann boundary condition for
 $\rho(r)$
\begin{equation}
\rho(0)\neq0,~~~~\dot{\rho}(0)=0,~~~~A(0)=0,~~~~f(0)=\pi.
\end{equation}
It is interesting to observe that the N-type solutions exhibit a non-vanishing
concentration at the core of vortex.

\subsection{\label{sec:level3} D-type Vortex Solutions}
% in LMH with Dirichlet Conditions}

We choose the following Dirichlet boundary conditions
\begin{equation}
\begin{aligned}
&\rho(0)=0,~~~~f(0)=\pi,~~~~A(0)=-\dfrac{1}{n},\\
&\rho(\infty)=2\sqrt{\dfrac{2\beta\mu-\alpha\gamma}{4\beta^2-\alpha^2}},~~~~
f(\infty)=\arccos{\dfrac{2\beta\gamma-\alpha\mu}{2\beta\mu-\alpha\gamma}},~~~~\\
&A(\infty)=\dfrac{2\beta(\gamma+\mu)-\alpha(\gamma+\mu)}{2(2\beta\mu-\alpha\gamma)}.
\label{eqn:Dboundary1}
\end{aligned}
\end{equation}
Numerical solutions of D-type vortices are obtained
for different values of $\gamma$.
We set the following parameter values
$g=1,~\alpha=0,~\beta=2,~\mu=1$, r is in the unit of $\sqrt{\beta/(2\mu)}$.

\begin{figure}
\includegraphics{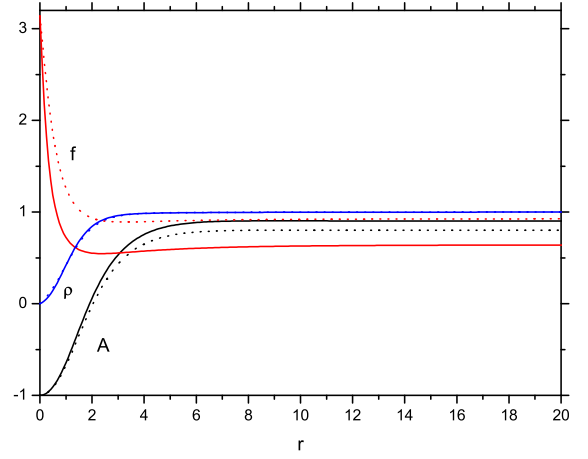}% Here is how to import EPS art
\caption{\label{fig:dvortex86} Solutions for $\rho,~f,~A$ with type-1.5 superconductivity with Dirichlet
boundary condition. We set
$n=1,~\alpha=0,~\beta=2,~\mu=1,~g=1$, and $r$ is given in units of $\sqrt{\beta/(2\mu)}$ in all figures.
Two solutions are shown for the cases $\gamma=0.8$ (solid lines) and
$\gamma=0.6$ (dotted lines).}
\end{figure}

As we mentioned, there exists type-1.5 and type-II vortex
with different mass ratios.
In this case we settle the parameters as $\gamma=0.8$ and $0.6$,
the respective mass ratios
$\beta_1=\sqrt{3.6},~\beta_2=\sqrt{0.4};~\beta_1=\sqrt{3.2},~\beta_2=\sqrt{0.8}$
correspond to type-1.5 superconductivity.
Solutions are shown in Fig.~\ref{fig:dvortex86}.

\begin{figure}
\includegraphics{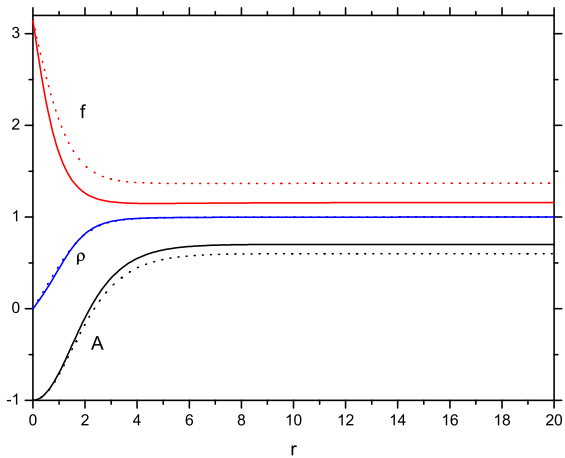}% Here is how to import EPS art
\caption{\label{fig:dvortex42} Solutions for $\rho,~f,~A$ with type-II superconductivity  corresponding
to Dirichlet boundary condition with $n=1,~\alpha=0,~\beta=2,~\mu=1,~g=1$.
Two solutions are shown for parameter values $\gamma=0.4$ (solid lines)
and $\gamma=0.2$ (dotted lines).}
\end{figure}

The magnetic flux is obtained as follows
\begin{equation}
\Phi=\oint{A_{\mu}dx^{\mu}}=[A(\infty)-A(0)]\frac{2\pi n}{g}.
\label{eq:flux}
\end{equation}
With Eq.~(\ref{eqn:Dboundary1}), the magnetic flux in Fig.~~\ref{fig:dvortex86} is
$n(0.9+\frac{1}{n})\Phi_0$ with $\gamma=0.8$, and $n(0.8+\frac{1}{n})\Phi_0$
with $\gamma=0.6$, where $~\Phi_0=\frac{2\pi}{g}~$ is the flux quanta.
It shows that the total magnetic flux through $x-y$ plane is fractional
multiple of the flux quanta.
Furthermore, a type-II superconductive phase appears with
$\gamma=0.4$ and $0.2$, which correspond to
$\beta_1=\sqrt{2.8},~\beta_2=\sqrt{1.2}$ and $\beta_1=\sqrt{2.4},~\beta_2=\sqrt{1.6}$.
Numerical solutions with different parameters are shown
in Fig.~\ref{fig:dvortex42}.
The total magnetic fluxes are $n(0.7+\frac{1}{n})\Phi_0$ and $n(0.6+\frac{1}{n})\Phi_0$
with $\gamma=0.4$ and $\gamma=0.2$, respectively.
The fluxes are shown to be fractional multiple of the flux quanta as well.

\subsection{\label{sec:level3} N-type Vortex Solutions} %in LMH with Neumann Conditions}
\begin{figure}
\includegraphics{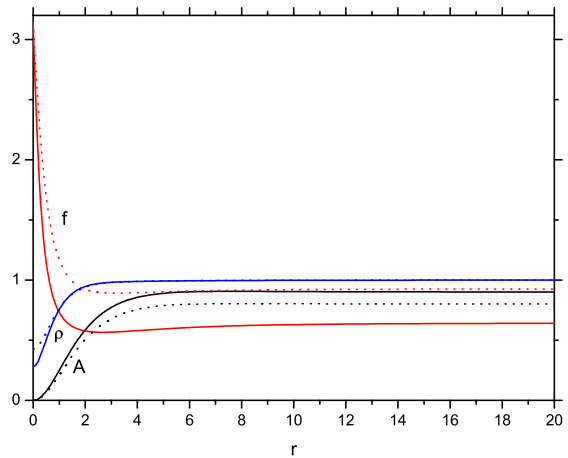}
\caption{\label{fig:nvortex86} Solutions for $\rho,~f,~A$ with type-1.5 superconductivity corresponding
to Neumann boundary condition with $n=1,\alpha=0,~\beta=2,~\mu=1,~g=1$.
Two solutions are shown for $\gamma=0.8$ (solid lines), $\gamma=0.6$ (dotted lines).}
\end{figure}

\begin{figure}
\includegraphics{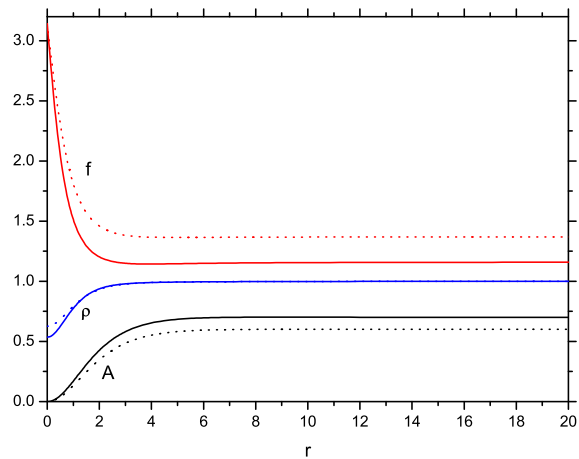}
\caption{\label{fig:nvortex42} Solutions for $\rho,~f,~A$ with type-II superconductivity
with Neumann boundary condition and with parameter values $n=1,~\alpha=0,~\beta=2,~\mu=1,~g=1$.
Two solutions are shown for $\gamma=0.4$ (solid lines), $\gamma=0.2$ (dotted lines).}
\end{figure}

In this case, we choose the following boundary conditions
\begin{equation}
\begin{aligned}
&\dot{\rho}(0)=0,~~~~f(0)=\pi,~~~~A(0)=0,\\
&\rho(\infty)=2\sqrt{\frac{2\beta\mu-\alpha\gamma}{4\beta^2-\alpha^2}},~~~~
f(\infty)=\arccos{\frac{2\beta\gamma-\alpha\mu}{2\beta\mu-\alpha\gamma}},~~~~\\
&A(\infty)=\frac{2\beta(\gamma+\mu)-\alpha(\gamma+\mu)}{2(2\beta\mu-\alpha\gamma)}.
\label{eq:Nboundary}
\end{aligned}
\end{equation}
Similarly, there are solutions with type-1.5 and type-II superconductivity.
We choose the same parameter $\gamma$ as we did in the case of D-type solutions.
Results of type-1.5 vortex with $\gamma=0.8,~0.6$ are shown in Fig.~\ref{fig:nvortex86},
and type-II vortex solutions with $\gamma=0.4,~0.2$ are shown in Fig.~\ref{fig:nvortex42}.

\begin{figure}
\includegraphics{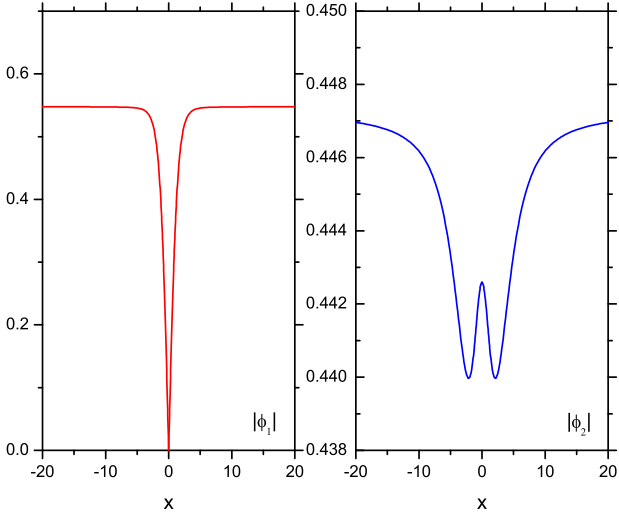}
\caption{\label{fig:rho12_sec} The behavior of condensates $\phi_1$ and $\phi_2$ in z-x plane
with $\alpha=0,~\beta=2,~\mu=1,c=0.2,~n=1,~g=1$ and $x$ is in the unit of $\sqrt{\beta/(2\mu)}$.
 $\phi_1$ component is shown to be the same as Abrikosov vortex.
Whereas, the $\phi_2$ component has a non-vanishing concentration at the core, the maximum  makes
the configuration of $\phi_2$ looks like W-shape.}
\end{figure}

\begin{figure}
\includegraphics{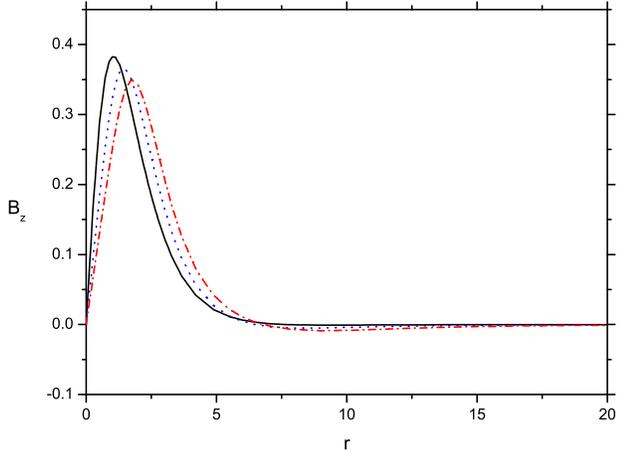}
\caption{\label{fig:Ax} The behavior of the magnetic field $B_z$ with
different winding numbers $n=1$ (solid line), $n=2$ (dotted line) and $n=3$ (dash-dotted line).
We choose the case of Neumann boundary conditions, and
$\alpha=0,~\beta=2,~\gamma=0.8,~\mu=1,~g=1$.}
\end{figure}

With Eq.~(\ref{eq:flux}) and Eq.~(\ref{eq:Nboundary}), the total magnetic
flux through the $x-y$ plane with $\gamma=0.8,~0.6,~0.4,~0.2$
are $\Phi=0.9n\Phi_0,~0.8n\Phi_0,~0.7n\Phi_0,~0.6n\Phi_0$, respectively. Then one can find that
N-type vortex has a magnetic flux smaller than D-type vortex has with the same parameters.
Although both two types of vortex carry infinite energy like the vortex of superfluid in GP theory,
one can always make a natural cut-off with a real size of the superconductor.
By this way one can obtain a finite energy vortex with fractional flux\cite{Cho08,Babaev02}.
Notice, the D-type vortex is topologically stable due to the presence
of non-trivial homotopy group $\Pi_2(S^2)$, and the
N-type vortex stability originates from the homotopy $\Pi_1(S^1)$.

The normal one-component superconductors have only integer
flux, whereas two-gap superconductors can possess both,
integer and fractional fluxes \cite{Babae04,Cho08,Babaev02}.
Furthermore, it can be found that the condensates concentration at core in Neumann type solution is non-zero $\rho(0)\neq0$.
The behavior of both two components $~|\phi_1|,~|\phi_2|~$ near
the core is described in Fig.~\ref{fig:rho12_sec} in $z-x$ plane.
Clearly, configuration of the component $~|\phi_1|~$ looks the same as the Abrikosov
vortex, while the behavior of $~|\phi_2|~$ is different,
the non-vanishing condensation at the core
makes it looks like the profile of $W$.
One should notice that such a behavior is caused by a specific electromagnetic
interaction of the two condensates irrespectively of interband
interaction\cite{Babaev09}.

There is another unexpected result relates to the magnetic
field properties depicted in Fig.~\ref{fig:Ax}.
The magnetic field reverses around at $r=7$ and then keeps
this opposite direction with a long decaying tail till
space infinity. According to numerical solution the inversion will be more
clear with a higher phase winding number $n$.

In Ref.~\cite{Babaev09} an unusual delocalization
of the magnetic field has been found which is
different from the normal Abrikosov vortex exponential localization.
This effect can take place in the superconductive phase of LMH as well
giving a new possible way to probe LMH at low temperature in experiment.

\section{\label{sec:level1}Vortex solutions with two-component complex phases}
We set both component condensates, $~\phi_1~$ and $~\phi_2~$, to have
nonzero complex phases.
The ansatz is chosen as follows
\begin{subequations}
\label{eq:ansatz2}
\begin{equation}
\begin{aligned}
\phi&
=\left(
\begin{array}{c}
\phi_1\\
\phi_2
\end{array}
\right)
=\dfrac{\rho(r)}{\sqrt{2}}
\left(
\begin{array}{c}
\cos{\dfrac{f(r)}{2}}\exp{(-in_1\varphi)}\\
\sin{\dfrac{f(r)}{2}}\exp{(-in_2\varphi)}
\end{array}
\right)
\\
&=\dfrac{\rho}{\sqrt{2}}
\exp{-ip\varphi}
\left(
\begin{array}{c}
\cos{\dfrac{f(r)}{2}}\exp{(-in\varphi)}\\
\sin{\dfrac{f(r)}{2}}
\end{array}
\right),
\end{aligned}
\end{equation}
\begin{equation}
A_{\mu}=\dfrac{q}{g}A(r)\partial_{\mu}\varphi.
\end{equation}
\end{subequations}
where $n=n_1-n_2$ and $p=n_2$.
With this, the Hamiltonian reads
\begin{equation}
\begin{aligned}
\mathcal{H}=&\dfrac{1}{2}\dot{\rho}^2+\dfrac{1}{8}\rho^2(\dot{f}^2
+\dfrac{n^2}{r^2}\sin^2{f})\\
&+\dfrac{\rho^2}{2r^2}(q A-n\dfrac{\cos{f}+1}{2}-p)^2
+\dfrac{q^2}{2g^2r^2}\dot{A}^2\\
&+\dfrac{\beta}{8}\big[(\rho^2-\dfrac{2\mu}{\beta})^2
+\dfrac{\alpha}{\beta}(\rho^2-\dfrac{4\gamma}{\alpha})\rho^2\cos{f}
+\rho^4\cos^2{f}\big]\\
&-\dfrac{\mu^2}{2\beta}.
\end{aligned}
\end{equation}
and equations of motion become
\begin{subequations}
\label{eqn:motion2}
\begin{equation}
\begin{aligned}
&\ddot{\rho}+\dfrac{1}{r}\dot{\rho}-[\dfrac{1}{4}(\dot{f}^2
+\dfrac{n^2}{r^2}\sin^2{f})\\
&+\dfrac{1}{r^2}(q A-n\dfrac{\cos{f}+1}{2}-p)^2]\rho
=\dfrac{\beta}{2}[(\rho^2-\dfrac{2\mu}{\lambda})\\
&+\dfrac{\alpha}{\beta}(\rho^2-\dfrac{2\gamma}{\alpha})\cos{f}
+\rho^2\cos^2{f}]\rho,
\end{aligned}
\end{equation}
\begin{equation}
\begin{aligned}
&\ddot{f}+(\dfrac{1}{r}+2\dfrac{\dot{\rho}}{\rho})\dot{f}
-2\dfrac{n}{r^2}(q A-\dfrac{n}{2}-p)\sin{f}\\
&=[2\gamma-(\dfrac{\alpha}{2}+\beta\cos{f})\rho^2]\sin{f},
\end{aligned}
\end{equation}
\begin{equation}
\ddot{A}-\dfrac{1}{r}\dot{A}-g^2\rho^2[A-\dfrac{n}{2q}(\cos{f}+1)-\dfrac{p}{q}]=0.
\end{equation}
\end{subequations}
The electromagnetic current is
\begin{equation}
j_{\mu}=g^2\rho^2(q A-n\dfrac{\cos{f}+1}{2}-p)\partial_{\mu}\varphi.
\label{eq:current2}
\end{equation}
Similarly, magnetic flux through the $~x-y~$ plane can be fixed as
\begin{equation}
\Phi=\oint{A_{\mu}dx^{\mu}}= [A(\infty)-A(0)]\dfrac{2\pi q}{g}.
\label{eq:flux2}
\end{equation}

The boundary condition for D-type and N-type vortices can be obtained
in a similar manner as in section 3. However, with two non-zero phase windings, there is
a third type, GP-type superconductivity due to chosen boundary condition.
In addition, one can find a neutral vortex which behaves as the vortex in the
superfluid which is described by Gross-Pitaevskii theory.

Let us consider the following boundary conditions:\\
The Dirichlet boundary conditions
\begin{equation}
\rho(0)=0,~f(0)=\pi,~A(0)=\dfrac{p\pm1}{q},
\end{equation}

The Neumann boundary conditions
\begin{equation}
\dot{\rho}(0)=0,~f(0)=\pi,~A(0)=\dfrac{p}{q},
\end{equation}

The Gross-Pitaevskii (GP) type boundary conditions for $n=2$
\begin{equation}
\rho(0)=0,~\dot{f}(0)=0,~A(0)=\dfrac{2p+n}{2q}.
\end{equation}

GP-type boundary conditions can be imposed also in the case of one component
phase winding for $n=2$, however, the corresponding vortex solution
is not stable. This is why we did not discuss it in section 3.
However, we can choose $n_1=1,~n_2=-1$, and $n=2$, then the composite vortices
are not only topologically but also thermodynamically stable\cite{Babae04,Babaev02}.
One can find that a neutral type vortex exists
not only with the Gross-Pitaevskii boundary condition,
but also with the Neumann boundary when $n=0$.

At infinity, with Eq.~(\ref{eq:vacuum}), Eq.~(\ref{eq:ansatz2})
and Eq.~(\ref{eq:current2}) one can obtain
\begin{equation}
\begin{aligned}
&\rho(\infty)=2\sqrt{\frac{2\beta\mu-\alpha\gamma}{4\beta^2-\alpha^2}},\\
&\cos{f(\infty)}=\frac{2\beta\gamma-\alpha\mu}{2\beta\mu-\alpha\gamma},\\
&A(\infty)=\frac{(2n+4p)\beta\mu-(n+2p)\alpha\gamma+2n\beta\gamma
-n\alpha\mu}{2q(2\beta\mu-\alpha\gamma)}.\\
\end{aligned}
\end{equation}
Then the total magnetic flux of the three types of vortex is
\begin{equation}
\Phi=\left\{
\begin{array}{ll}
\dfrac{n(\mu+\gamma)(2\beta-\alpha)+2(2\beta\mu-\alpha\gamma)}
{2q(2\beta\mu-\alpha\gamma)}\dfrac{2\pi q}{g} & \textrm{D-type}\\
\dfrac{n(\mu+\gamma)(2\beta-\alpha)}{2q(2\beta\mu-\alpha\gamma)}
\dfrac{2\pi q}{g} & \textrm{N-type}\\
\dfrac{2n\beta\gamma-n\alpha\mu}{2q(2\beta\mu-\alpha\gamma)}
\dfrac{2\pi q}{g} & \textrm{GP-type}
\end{array}
\right.
\end{equation}

In the case of presence of two complex phases there also
exist type-1.5 and type-II superconductivity. In the following,
we discuss two cases with particular winding numbers $n_1,~n_2$.

\subsection{$n_1=-n_2$}
\begin{figure}
\includegraphics{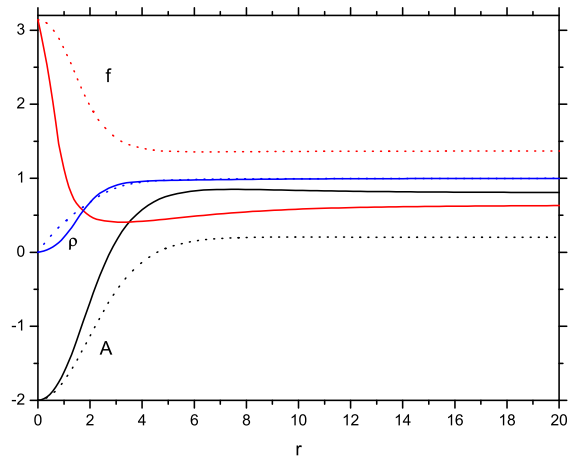}
\caption{\label{fig:n2p-1q1D} The D-type vortex solutions with $n=2,~p=-1,~q=1$.
Two solutions are shown with $\gamma=0.8$ (solid lines) $\gamma=0.2$ (dotted lines),
where $\alpha=0,~\beta=2,~\mu=1,~g=1$.}
\end{figure}

\begin{figure}
\includegraphics{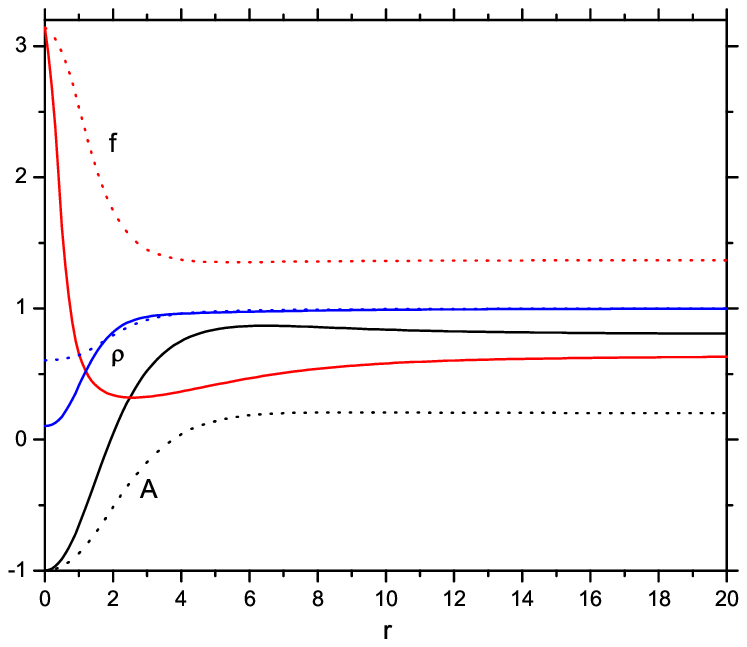}
\caption{\label{fig:n2p-1q1N} The N-type vortex solutions with $n=2,~p=-1,~q=1$.
Two solutions are shown with $\gamma=0.8$ (solid lines), $\gamma=0.2$ (dotted lines),
where $\alpha=0,~\beta=2,~\mu=1,~g=1$.}
\end{figure}

\begin{figure}
\includegraphics{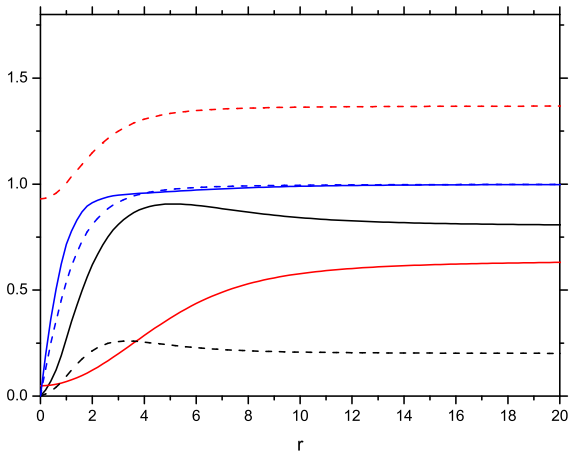}
\caption{\label{fig:GPvortex} Numerical solutions for the GP-type boundary conditions with $n=2,~p=-1,~q=1$, $\alpha=0,
~\beta=2,~\mu=1,~g=1$, $\gamma=0.8$ (solid lines), and $\gamma=0.2$ (dashed lines) and $r$.
The solutions for $A,~f,~\rho$ are depicted in black, red and blue respectively.}
\end{figure}

First, we consider a system with opposite phase windings $n_1=1,~n_2=-1$ (i.e.
$n=2,~p=-1,~q=1$). D-type vortex and N-type vortex
solutions are shown in Fig.~\ref{fig:n2p-1q1D} and Fig.~\ref{fig:n2p-1q1N}.
Moreover, there is a GP-type vortex with non-zero $f(0)$ in LMH,
non-trivial solutions are shown in Fig.~\ref{fig:GPvortex}
with $\gamma=0.8,~0.2$ and $\alpha=0,~\beta=2,~\mu=1,~g=1$.
In Fig.~\ref{fig:n2p-1q1D} and Fig.~\ref{fig:n2p-1q1N}
one can find that $f$ (dashed lines) decline monotonically from $\pi$,
while solid lines in these plots firstly decrease from $\pi$ to a minimum then increase to the vacuum expectation.
More interestingly, in Fig.~\ref{fig:GPvortex} lines $f$ of GP-type vortex increase monotonically near the core
which shows an opposite behavior comparing to the N-type and D-type vortex.
With the same parameters we choose for the three types of vortex,
one can find GP-type vortex carries the smallest flux.
So that, with an appropriate cut off,
one can more easily find GP-type vortex than N-type and D-type vortex in LMH in experiments.

\begin{figure}
\includegraphics{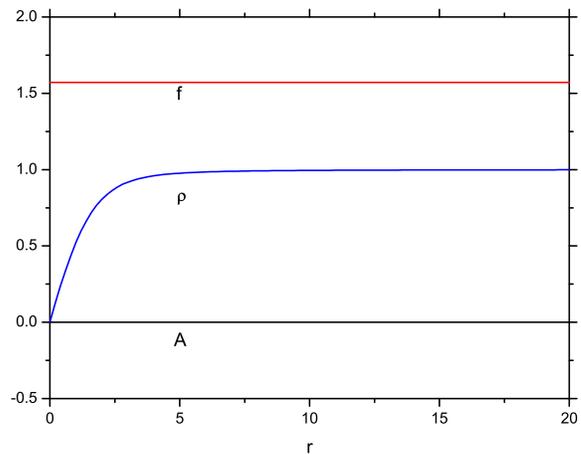}
\caption{\label{fig:GP} Numerical solutions for a neutral vortex with GP-type boundary conditions
with $n=2,~p=-1,~q=1$, $\alpha=0,
~\beta=2,~\mu=1,~g=1$, $\gamma=0$. The black, red and blue
lines correspond to solutions for $A,~f,~\rho$ respectively.
%It is a neutral
%vortex solution where $A\equiv0$ means that magnetic filed $B_z$ is zero.
}
\end{figure}

Moreover, it is noticed that there is a type of neutral
solution with GP-type boundary conditions in Eq.~(\ref{eqn:motion2})
\begin{equation}
\label{eq:neutral}
A\equiv 0,~~f\equiv\pi/2,~~\rho(0)=0
\end{equation}
We show it in Fig.~\ref{fig:GP}, a neutral type vortex exists in LMH with GP-type
boundary condition and
parameters (i.e. $n=2,~p=-1,~q=1$,$\alpha=0,
~\beta=2,~\mu=1,~g=1$, and $\gamma=0$). Without magnetic flux, this neutral vortex
looks like a vortex in superfluid described by the Gross-Pitaevskii theory.

One should notice, these results are consistent with results found in\cite{Babae04}.
It has been claimed that if the composite vortices ($\Delta \phi_1=2\pi,~\Delta \phi_2=-2\pi$)
in LMH were not yet ionized into two separated vortices, the both, superconductive superfluid phase and
metallic superfluid phase, can appear in LMH. Obviously, it is our case $n_1=1,~n_2=-1$.
We have demonstrated exactly that there is a super phase (metallic superfluid) with only neutral vortex
which is topologically and thermodynamically stable according to Babaev's arguments.
Solutions in Section 3 represent the case of $\Delta \phi_1=2\pi,~\Delta \phi_2=0$ in \cite{Babae04}.
These composite vortices are separated into two elemental vortices. Although,
they are not energetically favorable, they are topologically stable.

\subsection{The case of $n_1=n_2$ }
Unfortunately, this case is energetically forbidden, but this kind of vortex can be
induced by the vortex ($\Delta \phi_1=2\pi,~\Delta \phi_2=2\pi$) imposed in external field \cite{Babaev02}.
With $n_1=n_2$, naively, it seems that this two-component system is
identical to the ordinary one-component case, since these two components
lost their relative phase windings,
\begin{equation}
\phi=\dfrac{\rho(r)}{\sqrt{2}}
\left(
\begin{array}{c}
\cos{\dfrac{f(r)}{2}}\\
\sin{\dfrac{f(r)}{2}}
\end{array}
\right)\exp{(-in_1\varphi)}.
\end{equation}
This view is only partly correct with the GP-type boundary condition. One can check that
in such case it can be easily degenerated to the normal Abrikosov vortex in one
component superconductor with the condition
\begin{eqnarray}
f\equiv 0,~~A(0)=0,~~\rho(0)=0.
\end{eqnarray}

There is a more trivial solution
\begin{equation}
f(r) \equiv  const., ~~A(r) \equiv 1,~~\rho(0)=0.
\end{equation}
which is nothing but the neutral vortex in Gross-Pitaevskii theory.

\begin{figure}
\includegraphics{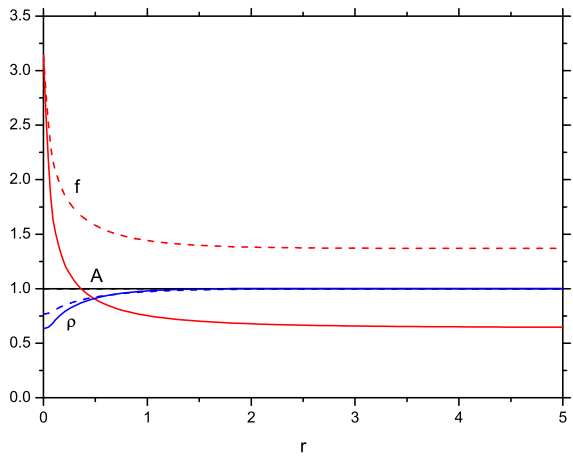}
\caption{\label{fig:np1q1} N-type neutral solutions with $n=0,~p=1,~q=1$,
and $\alpha=0,~\beta=2,~\mu=1,~g=1$.
Two neutral vortices are shown with $\gamma=0.8$ (solid lines) and
$\gamma=0.2$ (dashed lines).}
\end{figure}

Besides the above neutral vortex, we have found a non-trivial
neutral vortex with Neumann boundary condition. The neutral N-type vortex solutions with null magnetic flux
are shown in Fig.~\ref{fig:np1q1}. The magnetic field
cannot appear in the vortex in this metallic superconductor.
It indicates that there is a phase
corresponding to the magnetic superfluid state in which protonic Cooper pairs
coexist with the electronic Cooper pairs.

\section{Conclusion}
In this paper we have considered vortex solutions
with different topologies in various type superconductors.

We have found D-type, N-type and GP-type non-Abrikosov vortices according
to imposed different boundary conditions at the core.
We have shown that GP-type vortices carry a smaller magnetic
flux than N-type and D-type vortices.
The D-type vortex has no concentration of the condensate at the core,
whereas N-type vortex has a non-trivial profile of the condensate at the core.
In general, unlike the Abrikosov vortex,
the condensate $~\phi_2~$ in the N-type vortex in Fig.~\ref{fig:rho12_sec} has a local maximum
at the core which makes configuration looks like $W$,
while $\phi_1$ has the ordinary configuration as the Abrikosov vortex.
Furthermore, the magnetic flux in LMH can be integer only
in a special parameter limit, the fractional flux has been shown to be more general in LMH.
Another important property of non-Abrikosov vortex is the magnetic field inversion effect
which can be observed in the obtained delocalized solution in LMH.
It shows that vortex carries a positive magnetic field along
$Z$ axis near the center, while at a certain distance magnetic field flips to
the negative direction. This effect may also exist in the case when
both two condensates have non-zero complex phases as it is shown
in Fig.~\ref{fig:GPvortex} where the magnetic field is reversed clearly.

Moreover, there is a type of neutral vortex solution
with two non-zero complex phases. In this case, magnetic flux
cannot exist in the vortex, there is no super electromagnetic current
in LMH. This is similar to the vortex of superfluid in Gross-Pitaevskii theory.
Neutral vortex in LMH gives an important indication that there is a new
ordered state in LMH, the metallic superfluid\cite{Babae04}. That
new quantum ordered state cannot be purely categorized as a superconductor or
superfluid and deserves further study.

{\bf Acknowledgements}

Author wishes to express his gratitude to Professor Peng-ming Zhang
for multiple discussions and comments of the problem.
The work was supported by NSFC Grants (Nos. 11035006 and 11175215).



%% References


%%\bibliographystyle{model5-names}
%%\bibliography{vortex1.bib}

\begin{thebibliography}{00}
\bibitem{Babae04} E. Babaev, A. Sudb{\o}, N. W. Ashcroft, Nature 431 (2004) 666.
\bibitem{Cho08} Y. M. Cho , P. M. Zhang, Eur. Phys. J. B 65 (2008) 155.

\bibitem{Juan2012}  Juan C. Pi\~na,  Cl\'ecio C. de Souza Silva, and Milorad V. Milo\ifmmode \check{s}\else \v{s}\fi{}evi\ifmmode \acute{c}\else \'{c}\fi{}, Phys. Rev. B  86 (2012) 024512.

\bibitem{Babaev05} E. Babaev and M. Speight, Phys. Rev. B 72 (2005) 180502.

\bibitem{Geurt2010} R. Geurts, M. V. Milo\ifmmode \check{s}\else \v{s}\fi{}evi\ifmmode \acute{c}\else \'{c}\fi{}, and F. M. Peeters, Phys. Rev. B  81 (2010) 214514.

\bibitem{Chaves2011} A. Chaves, L. Komendov\'a, M. V. Milo\ifmmode \check{s}\else \v{s}\fi{}evi\ifmmode \acute{c}\else \'{c}\fi{}, J. S. Andrade. Jr, G. A. Farias, and F. M. Peeters,, Phys. Rev. B  83 (2011) 214523.

\bibitem{Babaev101} E. Babaev, J. Carlstrom, Physica C: Superconductivity 470 (2010) 717.
\bibitem{Babaev102} E. Babaev, J. Carlstrom, M. Speight, Phys. Rev. Lett. 105 (2010) 067003.
\bibitem{Johan1} J. Carlstrom, E. Babaev, M. Speight, Phys. Rev. B 83 (2011) 174509.
\bibitem{Johan2} J. Carlstrom, J. Garaud, E. Babaev, Phys. Rev. B 84 (2011) 134515.
\bibitem{Babaev2013} E. Babaev, M. Silaev, J. Supercond Nov Magn 26 (2013) 2045.


\bibitem{Moshchalkov09} V. Moshchalkov, M. Menghini, T. Nishio, Q. H. Chen, A. V. Silhanek, V. H. Dao, L. F. Chibotaru, N. D. Zhigadlo, and J. Karpinski, Phys. Rev. Lett. 102 (2009) 117001.


\bibitem{vagov2012} A. Vagov, A. A. Shanenko, M. V. Milo\ifmmode \check{s}\else \v{s}\fi{}evi\ifmmode \acute{c}\else \'{c}\fi{}, V. M. Axt, and F. M. Peeters, Phys. Rev. B  86 (2012) 144514.
\bibitem{Orlova2013} N. V. Orlova, A. A. Shanenko, M. V. Milo\ifmmode \check{s}\else \v{s}\fi{}evi\ifmmode \acute{c}\else \'{c}\fi{}, and F. M. Peeters, Phys. Rev. B 87  (2013) 134510.

\bibitem{Ashcroft68} N. W. Ashcroft, Phys. Rev. Lett. 21 (1968) 1748.
\bibitem{Moulopolos91} K. Moulopolos and N. W. Ashcroft, Phys. Rev. Lett. 66 (1991) 2915.
\bibitem{Babaev02} E. Babaev, Phys. Rev. Lett. 89 (2002) 067001.
\bibitem{Babaev09} E. Babaev, J. J\"aykk\"a, M. Speight, Phys. Rev. Lett. 103 (2009) 237002.
\bibitem{Forgacs06} P. Forg\'acs, S. Reuillon, M. S. Volkov, Phys. Rev. Lett. 96 (2006) 041601.
\bibitem{Zhitomirsky04} M. E. Zhitomirsky, V. H.Dao, Phys. Rev. B 69 (2004) 054508.
\bibitem{Hindmarsh92} M. Hindmarsh, Phys. Rev. Lett. 68 (1992) 1263.
\bibitem{Nielsen73} H. B. Nielsen, P. Olesen, Nucl. Phys. B 61 (1973) 45.
\bibitem{Cho05PRB} Y. M. Cho Phys. Rev. B 72 (2005) 212516.
\bibitem{Horvathy2009} P. A. Horvathy, P. M. Zhang, Phys. Rept. 481 (2009) 83.
\end{thebibliography}



\end{document}